\documentclass{article}

\usepackage{authblk}
\usepackage{amsmath,amssymb}
\usepackage{bm}
\usepackage{graphicx}
\usepackage{array}
\usepackage{amssymb,dsfont,mathrsfs}
\usepackage[mathscr]{eucal}
\usepackage{subfigure}
\usepackage{verbatim}
\usepackage{wrapfig}
\usepackage{cite}
\usepackage{makeidx}
\usepackage{enumerate}
\usepackage{empheq}
\usepackage{float}
\usepackage{threeparttable}
\usepackage{caption}
\usepackage[left=25truemm,right=25truemm]{geometry}
\usepackage{txfonts}
\usepackage{bm}
\usepackage{threeparttable}

\title{Notes on the Fast Multipole Method}

\author{Yasuhiro Kajima\thanks{y-kajima@nifty.com, \ kajima@nzu.ac.jp}}
\affil{Nagoya Zokei University, 2-4-1 Meijo, Kitaku, Nagoya-shi, Aichi, 462-0846, Japan}
\begin{document}
\maketitle

\begin{abstract}
{Coulomb interactions of point charges can be calculated in $\mathcal{O}$(N) computation using the fast multipole
method and direct calculations between charges nearby.  It reduces computational cost dramatically, however, because of
its method that combines direct and indirect calculations, there exists discontinuity of potential energy
 with respect to positions of charges. 
In this paper, we remove Legendre functions usually used in the fast multipole method 
and instead use charges fixed in positions.
As an application of this method, we remove the discontinuity. 
It also leads us to a method of periodic boundary condition that is 
continuous even if a particle goes out from a wall of a simulation box and enters in opposite side of the box. 
Lastly, we show a version of the fast multipole method that do not use shift process.} 
\end{abstract}

\section{Introduction}
Because the fast multipole method (FMM) developed by Greengard and Rokhlin \cite{green} makes it possible 
to compute $n$-body problems 
in $\mathcal{O}(N)$ calculation with predictable error bounds, it has been applied to many fields 
that require much computation time such as the density functional theory,\cite{Ramzi} 
as well as molecular dynamics simulation.
It is considered as one of the top 10 algorithms of the 20th century. \cite{110}
The original FMM includes computing interactions of charges in 3-dimensional space and it has 
become an important tool for performing molecular simulation. 
We briefly recall the basic idea of FMM in a simplified form necessary to describe our method.
We assume a simulation box is a cube, and the number of point charges are $N$.

In molecular simulation, potential energy or electrostatic forces are computed as the sum of pairwise interactions
of charges in a simulation box.
To apply FMM to the computation of such physical quantities, we first divide the simulation box
into small cells (cubes) by dividing each edge into
$\frac{1}{2^k}$ in size for $0\le k\le n$.  
If a cell is obtained in the $k$-th division, we call the cell is level $k$ and
if $k=n$, we call the cell a leaf cell or a finest cell.
Then, every  cell of $k$-th level ($k<n$) is composed of eight smaller cells of $(k+1)$-th level.
Interactions of every two charges each in well-separated cells\cite{green}
are computed and summed using FMM technique to obtain {\it far-field} interactions.
The rest pairwise interactions due to charges in the same finest cell or adjacent 26 finest cells
 ({\it near-field} interaction, we do not include second nearest neighbor) are computed directly.
 The sum of these near- and far-field interactions gives us the total potential or electrostatic forces.
This computation requires $\mathcal{O}(N)$ operations.
The combination of FMM and direct calculation enables us to perform 
large-scale simulations \cite{140} within an affordable amount of time.
However, since FMM treats each particle in a different way according to which leaf cell it belongs, 
the total potential and the force a particle feels change discontinuously when it goes across 
boundaries of leaf cells. It implies that a particle may move abnormally near the boundaries.

In the next section, we introduce a version of FMM and
a method where potential and forces are continuous even if point charges go across boundaries.
It can be applied to 
the periodic boundary conditions to make it continuous even when a charged particle goes out of the outermost boundary
of the simulation box. 
We investigate accuracies of our method in \S 3.
We discuss the results obtained in the previous section in \S\ref{diss}.
We also investigate the errors of our method, and
another application of our method that can 
remove "shift" process from our FMM in the Appendix\ref{appBi}. 

\section{Theory}
\subsection{Replacement, shift, and its invariance property}
In this section, we introduce notions of {\it replacement}, {\it shift}, and their {\it invariance property}.
A {\it replacement} is a linear mapping from point charges to a vector, a {\it shift} is a linear mapping
from a vector to a vector, and its {\it invariance property} is a relationship between compositions of these mappings.
First, we introduce some tools.
\subsubsection{Lagrange interpolation}
First, we recall Lagrange interpolation; for a polynomial $f(x)$ of deg$(f) < n$ we have
\begin{equation}
\label{Lag1}
\sum^n_{i=1} f(a_i)g_i(x) = f(x) 
\end{equation}
where $a_i$'s are $n$ distinct numbers and $g_i(x):=\frac{\prod_{k\ne i}(x-a_k)}{ \prod_{k\ne i}(a_i-a_k)}$.
Note that $g_i(a_j)=\delta_{ij}$ (Kronecker's delta function) and
(\ref{Lag1}) is equivalent to 
\begin{equation}
\label{Lag2}
\sum^n_{i=1} a_i^j g_i(x) = x^j
\end{equation}
for all $j< n$.

\subsubsection{$n$-division points on a segment}
In a space with a coordinate system, we denote by 
$S^x$ a segment parallel to the $x$-axis, and by $S_n^x$
the equally spaced $n$ points on $S^x$ including the end points of $S^x$.
We call $S_n^x$ the {\it n-division points} of $S^x$.
We also call the $x$-coordinate of $S_n^x$ as {\it n-division points}
and denote it by the same symbol $S^x_n$ if there is no risk of confusion.
A case where $n=4,$ the length of $S^x$$=6$, and the origin is the middle point of $S^x$, 
is illustrated in Fig.1.
We define similarly for $y$-axis and $z$-axis.

\begin{figure}
\begin{center}
\includegraphics[width=8cm]{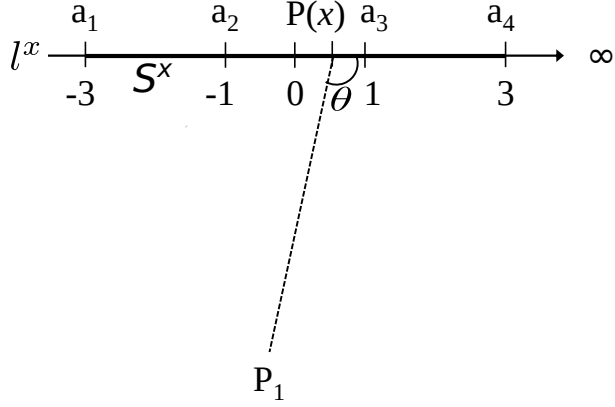}
\caption{{\it $n$-division points} on a segment $S^x$.
A segment $S^x$ and its {\it n-division points} $S_n^x$ lie on a line $l^x$.
$P(x)$ is a point on the line whose coordinate is $x$.
$P_1$ is a fixed point different from $P(0)$ or $P(x)$.
An angle $\theta$ and a length $r$ are defined by $\theta:=P_1P(x)P(\infty)$ and $r:=P(0)P_1$.
The figure is depicted for the case $n=4$,
the length of $S^x=6$, and the origin is the middle point of $S^x$.
}
\label{F1}
\end{center}
\end{figure}

\subsubsection{Coulomb potential and one-dimensional replacement}
\label{Coulomb_one}
Take the notations as before.
We take the middle point of $S^x$ as origin and denote by $O$.
Let $l^x$ be a line containing $S^x$,
$P(x)$ be a point on the line whose $x$-coordinate is $x$
(not necessarily on $S^x$), and
$P_1$ be an arbitrarily fixed point different from $P(x)$ or $O$.
Put $F(P(x),P_1):=\frac{1}{|P(x)-P_1|}$.  
Putting $r$ the distance from $P_1$ to $O$, 
we have\cite{green}  $F(P(x),P_1)=\sum^\infty_{j=0} L_j(u){x^j}{r^{-(j+1)}}$ for $|x/r|<1$ where $L_j(u)$'s are Legendre polynomials,
$u=\cos\theta$, and $\theta$ 
is the angle $P_1P(x)P(\infty)$ as depicted in Fig.1.
Then, for $a_i\in S^x_n$, we have
$F(P(x),P_1) -\sum^n_{i=1} F(P(a_i),P_1)g_i(x) = \sum^\infty_{j=0} L_j(u)({x^j - \sum^n_{i=1} a_i^j g_i(x)}){r^{-(j+1)}} 
= \sum^\infty_{j=n} L_j(u)(x^j- \sum^n_{i=1}  a_i^j g_i(x)){r^{-(j+1)}} $
where we used (\ref{Lag2}).
Therefore, we can use $\sum^n_{i=1} F(P(a_i),P_1)g_i(x)$ for $F(P(x),P_1)$
with the error $E_r^x(x):=\sum^\infty_{j=n} L_j(u)(x^j- \sum^n_{i=1}  a_i^j g_i(x)){r^{-(j+1)}} $,
\begin{equation}
\label{error}
F(P(x),P_1) =\sum^n_{i=1} F(P(a_i),P_1)g_i(x) +E^x_r(x).
\end{equation}
Denoting the order of $E_r^x(x)$ with respect to $r$ by $\mathcal{O}_r(E_r^x(x))$, we find $\mathcal{O}_r(E_r^x(x))\leq-(n+1)$. 
We discuss the error in more detail in the Appendix.

We denote a point charge at $P(x)$ on $l^x$ whose charge strength is $q$ by $(P(x), q)$.
Multiplying both sides of (\ref{error}) by $q$, we find
the Coulomb potential arises from this charge at $P_1$ is equal to 
the potential arises from point charges  $\{(P(a_i), qg_i(x)) \  \mid a_i\in S^x_n, 1\leq i\leq n \}$
with an error $qE^x_r(x)$.
We call the mapping from a point charge to a $n$-dimensional vector
\begin{equation}
Rpl _{S_n^x} : (P(x), q) \longrightarrow (qg_i(x)) \ (1\le i\le n)
\end{equation}
as {\it replacement} of order $n$ with respect to $S_n^x$.
If there are multiple point charges on the line $l^x$, we extend the mapping $Rpl _{S_n^x}$ by linearity.
We note the {\it replacement} does not depend on the choice of the $x$-coordinate.
Neither translating nor scaling the $x$-axis have effect on the replacement.
It is determined geometrically.

\subsubsection{$n$-division points of a cube}
\label{ndiv3}
\begin{figure}
\begin{center}
\includegraphics[width=8cm]{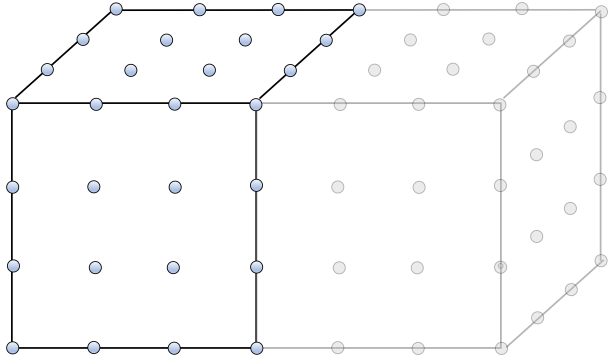}
\caption{{\it n-division points} for a cube (left).
A cube and its $n$-division points are depicted for the case $n=4$.
A simulation box is divided into cells, and 
two adjacent cells of the same level have some
$n$-division points on the same positions.
We treat separately, that is, treat as different point charges on the same position
(\S\ref{rep3c}).}
\label{F2}
\end{center}
\end{figure}

As before, we assume a simulation box is a cube for simplicity.
We extend the $n$-{\it division points} of a segment also to a cube.
Suppose we are given a number $n$ and a cube $C$ whose edges $C^x$, $C^y$, and $C^z$ 
are parallel to the coordinate axes. 
(The choice of the edges does not affect the following.)
Then, we define $C^x_n$, $C^y_n$, and $C^z_n$ as the {\it $n$-division points} of $C^x$, $C^y$, and $C^z$, respectively.
We denote by $C_n$ the lattice points 
$\{(a_i^x, a_j^y, a_k^z)\mid a_i^x\in C^x_n, a_j^y\in C^y_n, a_k^z\in C^z_n \}$
in $C$.
These are {\it n-division points} of the cube $C$.
Figure \ref{F2} shows the {\it n-division points} for $n=4$.
We also denote a point charge $(P,q)$ whose coordinate is
$(x,y,z)$ by $(P(x,y,z),q)$.
We use $P(a_{i,j,k})$ as a shorthand notation for $P(a_i^x,a_j^y,a_k^z)$.

\subsubsection{Replacement for three-dimensional cases}
\label{rep3c}
Let $C$ be a cube defined by $\{(u,v,w)\mid x_0\le u\le x_1, y_0\le v\le y_1, z_0\le w\le z_1\}$
and $(P(x,y,z),q)$ be a point charge (not necessarily in $C$).
We draw a line $l^x$ parallel to the $x$-axis through $P(x,y,z)$.
We identify the
segment $P(u), (\ x_0\le u\le x_1)$ on $l^x$ with $C^x$.
Apply one-dimensional {\it replacement}  with respect to $C^x_n$ to the point charge $(P,q)$,
then we get $\{(P(a_i^x,y,z), qg_i(x)) \mid 1\leq i\leq n \}$.
Next, we apply one-dimensional replacement with respect to $C^y_n$ to each of these $n$ point charges,
we get $\{(P(a_i^x,a_j^y,z), qg_i(x)g_j(y)) \mid 1\leq i, j\leq n \}$.
Finally, apply the replacement with respect to $C^z_n$ to each of these $n^2$ point charges.
Then, we get $n^3$ point charges
$\{(P(a_i^x,a_j^y,a_k^z), qg_i(x)g_j(y)g_k(z)) \mid 1\leq i, j, k\leq n \}$.
Put $g_{i,j,k}^C(x,y,z):=g_i(x)g_j(y)g_k(z)$.
The superscript $C$ is used to specify the cube $C$.
Then the mapping
\begin{equation}
\label{exp1}
(P,q) \rightarrow  (qg_{i,j,k}^C(x,y,z)) \ (1\leq i, j, k\leq n)
\end{equation}
maps a point charge $(P,q)$
to a $n^3$-dimensional vector whose entries are charge strengths of 
{\it $n$-division points} $P(a_{i,j,k})$ $(a_{i,j,k}\in C_n$).
When there are multiple point charges, we extend the mapping by linearity.
In other words, $m$ point charges $Pc:=\{(P(x_l,y_l,z_l),q_l)\mid 1\leq l\leq m\}$
is mapped to $\sum_{1\leq l\leq m}q_lg_{i,j,k}^C(x_l,y_l,z_l)$.
We call the linear mapping as a {\it replacement} of order $n$ with respect to $C_n$ and denote as follows:
\begin{equation}
\label{ex66}
Rpl_{C_n}: Pc \rightarrow \Bigl(
\sum_{1\leq l\leq m}q_lg_{i,j,k}^C(x_l,y_l,z_l)\Bigr)  \ (1\leq i, j, k\leq n).
\end{equation}
The $n^3$ point charges on $C_n$ take the place of $Pc$.
Since $g_{i,j,k}^C(x,y,z)=g_i(x)g_j(y)g_k(z)$ is a symmetric product, 
this mapping, which is a successive operations of three {\it replacements} of one-dimension,
does not depend on the order (permutations) of the operations.  
It is determined geometrically by their relative positions of the cube and the point charges.
The {\it replacement} (\ref{ex66}) plays a similar role to the {\it forming multipole expansions}. \cite{green}

\subsubsection{Shift and invariance property}
Next, we introduce {\it shift} and {\it invariance property} of the {\it replacement}. 
First, we define {\it shift} for a one-dimensional case.
We assume that two segments $S^x$ and $\tilde{S}^x$ are given on a line $l^x$.
For $a_i\in S^x_n$  and $\tilde{a}_i\in \tilde{S}^x_n$, we define
$n$-th degree polynomials $g_i(x)$ and $\tilde{g}_i(x)$ by 
$g_i(x):=\frac{\prod_{k\ne i}(x-a_k)}{ \prod_{k\ne i}(a_i-a_k)}$ and
$\tilde{g}_i(x):=\frac{\prod_{k\ne i}(x-\tilde{a_k})}{ \prod_{k\ne i}(\tilde{a_i}-\tilde{a_k})}$, respectively.
Since the degree of $\tilde{g}_i(x)$ is $n-1$, we have
$\sum_{i=1}^n \tilde{g}_j(a_i)g_i(x)=\tilde{g}_j(x)$ from (\ref{Lag1}). Thus we have
\begin{equation}
\label{trans}
\sum_{i=1}^n q\tilde{g}_j(a_i)g_i(x)=q\tilde{g}_j(x)
\end{equation}
for $1\leq j \leq n$.

Now we define a linear transformation 
$\mathit{shift}_{S^x_n\rightarrow \tilde{S}^x_n} : (q_1, q_2,\cdots,q_n ) \longrightarrow 
(\tilde{q}_1, \tilde{q}_2,\cdots,\tilde{q}_n )$
by
\begin{equation}
(\tilde{q}_1, \tilde{q}_2,\cdots,\tilde{q}_n )^{\rm T}=
\overleftrightarrow{G}_{S^x_n\rightarrow \tilde{S}^x_n}(q_1, q_2,\cdots,q_n )^{\rm T},
\end{equation}
where $\overleftrightarrow{G}_{S^x_n\rightarrow \tilde{S}^x_n}$ 
is a matrix whose $(j,i)$-entry is $\tilde{g}_j(a_i)$ and the superscript ${\rm T}$
represents the transpose operation.

A point charge $(P(x),q)$ is {\it replaced} to a vector $(qg_i(x))$ $(1\leq i\leq n)$
with respect to $S_n^x$.
Then, by $\mathit{shift}_{S^x_n\rightarrow \tilde{S}^x_n}$, 
the vector $(qg_i(x))$
is transformed to a vector $(\tilde{q}_1, \tilde{q}_2,\cdots,\tilde{q}_n )$
where $\tilde{q}_j=\sum_i q\tilde{g}_j(a_i)g_i(x)=q\tilde{g}_j(x)$ by (\ref{trans}).
Thus we have $(\tilde{q}_1, \tilde{q}_2,\cdots,\tilde{q}_n )=(q\tilde{g}_1(x), q\tilde{g}_2(x),\cdots,q\tilde{g}_n(x) )$,
which implies
\begin{equation}
\label{onediminv}
Rpl_{\tilde{S}_n^x}=\mathit{shift}_{S_n^x\rightarrow \tilde{S}_n^x}\circ Rpl_{S_n^x}.
\end{equation}
We call this property as {\it invariance property}.

This {\it shift} is extended to three-dimensional cases, and the
 invariance property also holds.
Let $C$ and $\tilde{C}$ be cubes.
With notation as before, 
${\it shift}_{C_n\rightarrow \tilde{C}_n}:(q_{i,j,k})\rightarrow (\tilde{q}_{i,j,k})$ 
is defined 
as the composition of the following three {\it shifts}:
\[
(q_{1,j,k}^x, q_{2,j,k}^x,\cdots,q_{n,j,k}^x)^{\rm T}=
\overleftrightarrow{G}_{C^x_n\rightarrow \tilde{C}^x_n}(q_{1,j,k}, q_{2,j,k},\cdots,q_{n,j,k} )^{\rm T},
\]
\begin{equation}
\label{threethree}
(q_{i,1,k}^{x,y}, q_{i,2,k}^{x,y},\cdots,q_{i,n,k}^{x,y})^{\rm T}=
\overleftrightarrow{G}_{C^y_n\rightarrow \tilde{C}^y_n}
(q_{i,1,k}^{x}, q_{i,2,k}^{x},\cdots,q_{i,n,k}^{x})^{\rm T},
\end{equation}
\[
(\tilde{q}_{i,j,1}, \tilde{q}_{i,j,2}\cdots,\tilde{q}_{i,j,n})^{\rm T}=
\overleftrightarrow{G}_{C^z_n\rightarrow \tilde{C}^z_n}
(q_{i,j,1}^{x,y}, q_{i,j,2}^{x,y},\cdots,q_{i,j,n}^{x,y})^{\rm T}.
\]
Each {\it shift} transforms $n^2$ vectors of $n$-dimension.
We denote this transform by $\mathit{shift}_{C_n\rightarrow \tilde{C}_n}$.
From the construction (\ref{threethree}), the {\it shift} is invariant under permutations 
of the compositions.
Due to the fact that {\it replacement} and {\it shift} are 
both composed of three operations each operate only on one axis,
the {\it invariance property} for one-dimensional case (\ref{onediminv})
leads that for a three-dimensional case below.
\begin{equation}
Rpl_{\tilde{C}_n}=\mathit{shift}_{C_n\rightarrow \tilde{C}_n}\circ Rpl_{C_n}.
\end{equation}
We abbreviate the $\mathit{shift}_{C_n\rightarrow \tilde{C}_n}$ as $\mathit{shift}_{C\rightarrow \tilde{C}}$ 
when $n$ is clear from the context.
The $\mathit{shift}_{C\rightarrow \tilde{C}}$ is also a linear transformation
of $n^3$-dimensional vector space.

\subsubsection{Multipole to local expansion}
\label{expansions}
We can derive a representation similar to the {\it multipole to local expansion}.
Suppose we are given two disjoint cubes $C$, $\tilde{C}$ and
point charges $(P,q)\in C$, $(\tilde{P},\tilde{q})\in\tilde{C}$, and a number $n$. 
Then, similar to (\ref{error}) we have
\begin{equation}
\label{newer}
q\tilde{q}F(P,\tilde{P})
=\sum_{i,j,k}\sum_{\tilde{i},\tilde{j},\tilde{k}}g_{i,j,k}^{C}
g_{\tilde{i},\tilde{j},\tilde{k}}^{\tilde{C}}F(P(a_{i,j,k}),P(\tilde{a}_{\tilde{i},\tilde{j},\tilde{k}})),
\end{equation}
where we have omitted the error term.
For multiple point charges, we define
$Pc:=\{(P(x_l,y_l,z_l),q_l)\mid 1\leq l\leq m\}$ the point charges in a cell $C$, and
$\mathfrak{g}^{C}=(\mathfrak{g}_{i,j,k}^{C}):=(\sum_{1\leq l\leq m}q_lg_{i,j,k}^C(x_l,y_l,z_l))$ 
for the cell $C$.
Then we have
\begin{equation}
\label{fmm}
\sum_{(P,q)\in Pc}\sum_{(\tilde{P},\tilde{q})\in P\tilde{c}}q\tilde{q}F(P,\tilde{P})
=\sum_{i,j,k}\sum_{\tilde{i},\tilde{j},\tilde{k}}\mathfrak{g}_{i,j,k}^{C}
\mathfrak{g}_{\tilde{i},\tilde{j},\tilde{k}}^{\tilde{C}}F(P(a_{i,j,k}),P(\tilde{a}_{\tilde{i},\tilde{j},\tilde{k}}))
\end{equation}
where we have omitted the error term.
Since a three-dimensional {\it replacement} is an iterative application of one-dimensional {\it replacements},
the order of omitted error with respect to $r$ is also $\le-(n+1)$. 
This error term will be discussed in the appendix.
The equation (\ref{fmm}) allows us to compute a sum of pairwise potentials between point charges in
$C$ and $\tilde{C}$ by computing those between $C_n$ and $\tilde{C}_n$.
The equation plays a similar role to {\it multipole to local expansions}.

In addition, since the two vectors
$\mathfrak{g}^{C}$ and 
$\mathfrak{g}^{\Tilde{C}}$
are both $n^3$-dimensional vectors, we can write
the terms 
\[
\sum_{i,j,k}\sum_{\tilde{i},\tilde{j},\tilde{k}}\mathfrak{g}_{i,j,k}^{C}
\mathfrak{g}_{\tilde{i},\tilde{j},\tilde{k}}^{\tilde{C}}F(P(a_{i,j,k}),P(\tilde{a}_{\tilde{i},\tilde{j},\tilde{k}}))
\]
in the right-hand side of (\ref{fmm}) 
as $B_{C,\tilde{C}}^n(\mathfrak{g}^{C},\mathfrak{g}^{\Tilde{C}})$,
where we put
\begin{equation}
\label{bilinear}
B_{C,\tilde{C}}^n(\bm{v},\bm{w}):=
\sum_{i,j,k}\sum_{\tilde{i},\tilde{j},\tilde{k}}v_{i,j,k}
w_{\tilde{i},\tilde{j},\tilde{k}}F(P(a_{i,j,k}),P(\tilde{a}_{\tilde{i},\tilde{j},\tilde{k}})).
\end{equation}
Here $B_{C,\tilde{C}}^n(\bm{v},\bm{w})$ is a bilinear form with respect to
$\bm{v}$ and $\bm{w}$, which are indexed as $\bm{v}=(v_{i,j,k})$, $\bm{w}=(w_{i,j,k})$, respectively.

\subsection{Formulation of FMM}
\label{formulation}
The FMM based on our method proceeds
in the following steps.\cite{green,ogata} We assume the simulation box has a cubic shape
whose dimension is $(h, h, h)$. There are point charges in the simulation box and
the {\it order} of {\it replacement} is $n$.
\begin{enumerate}
\item {\it Divide the simulation box.} 
We divide the simulation box into cells with dimensions
$(h/2^{l}, h/2^{l}, h/2^{l})$ for level $l$ ($1\le l\le l_f$). 
If $l_f=0$ or $=1$, there are no well-separated cells. Thus we assume $l_f\geq 2$.
If a cell $C$ is of level $l$, we indicate it by adding a subscript $(l)$ as $C_{(l)}$
if necessary.

\item {\it Upward Pass.}
For each finest cell $C^i_{(l_f)}$, {\it replace} the all point charges in the cell 
with respect to it to obtain an $n^3$-dimensional vector
$\Phi_{C_{(l_f)}^i}$, ($1\le i\le 8^{l_f}$).
Then, we {\it shift} the vectors $\Phi_{C_{(l_f)}^i}$ to their parent cells $C_{(l_f-1)}^j$
and add the {\it shifted} eight vectors for each $C_{(l_f-1)}^j$. 
We denote the resulting vector by $\Phi_{C_{(l_f-1)}^j}$.
Iterating these procedures upward, we obtain $\Phi_{C_{(l)}^i}$ for all $l\le l_f$ and $i$.
$\Phi_{C_{(l)}^i}$ is equal to the vector obtained by replacing the point charges in $C_{(l)}^i$
with respect to $C_{(l)}^i$ (thus equal to $\mathfrak{g}^{C_{(l)}^i})$.

\item {\it Downward Pass.}
For a cell $C$, we define linear functions
$\Psi_C(\bm{v})$ from an $n^3$-dimensional vector $\bm{v}$ to a real number
as follows:
First, put $\Psi_{C^0_{(0)}}(\bm{v})=\Psi_{C^i_{(1)}}(\bm{v})=0$ for all $1\leq i\leq 8^1$.
Second, suppose $\Psi_{C^i_{(l-1)}}(\bm{v}')$ have been obtained for some $l$ $(\ge2)$ and all $1\le i\le 8^{l-1}$.
For a child $C_{(l)}^j$ of $C^i_{(l-1)}$ denote the function
$\Psi_{C^i_{(l-1)}}(\mathit{shift}_{C^j_{(l)}\rightarrow C^i_{(l-1)}}(\bm{v}))$
by $\Psi_{C^j_{(l)}}(\bm{v})$.
Third, add
$B^n_{C^j_{(l)},C^k_{(l)}}(\bm{v},\Phi_{C_{(l)}^k})$
to $\Psi_{C^j_{(l)}}(\bm{v})$ for all
$C^k_{(l)}$ in the interaction list\cite{green} of $C^j_{(l)}$.
We again denote the resulting function by $\Psi_{C^j_{(l)}}(\bm{v})$.
Iterating the procedures downward, we obtain $\Psi_{C^j_{(l_f)}}(\bm{v})$ for all $j$.

\item
{\it Far-field potential.}
Once we have obtained $\Psi_{C^j_{(l_f)}}(\bm{v})$,
$\Psi_{C^j_{(l_f)}}(\Phi_{C_{(l_f)}^j})$ gives us the potential between charges in $C^{j}_{(l_f)}$ and 
charges not in $C^{j}_{(l_f)}$ and not in its nearest neighbors. 
Adding $\Psi_{C^j_{(l_f)}}(\Phi_{C_{(l_f)}^j})$ for all $1\le j\le 8^{l_f}$, we obtain total far-field potential.

\item
{\it Total potential.}
For each point charge in a finest cell $C^{j}_{(l_f)}$, we
compute directly the potentials between the point charge and point charges in the cell $C^{j}_{(l_f)}$ itself
and  its nearest neighbors.
Add all these values and the value obtained in (4) together, 
we obtain the total potential.
\end{enumerate}
We can compute the force a particle $P=(P(x_1,y_1,z_1),q) \in C_{(l_f)}$
feels similar to the method above.
We compute the force from near-field directly.
The far-field part is computed as follows.
First, we compute $\bm{v'}_a=-q\frac{\partial}{\partial a}g^C_{i,j,k}(x,y,z)|_{(x,y,z)=(x_1,y_1,z_1)}$ for $a\in \{x,y,z\}$
(see (\ref{exp1})).
Then the force $P$ feels due to the far-field force is = $\Psi_{C^j_{(l_f)}}(\bm{v'}_a)$.
Add both values from near-field and far-field together, then it gives us the force.

\subsection{Continuation on boundaries}
\label{contBoun}
In a molecular dynamics simulation, point charges (charged particles) 
may go beyond the borders of 
the finest cells.
The moment a particle goes across the borders, its electrostatic potential 
is computed by different equations in the framework of FMM, which can cause discontinuity of potential energy
and force, and thus abnormal behavior of particles.
 
\begin{figure}
\begin{center}
\includegraphics[width=8cm]{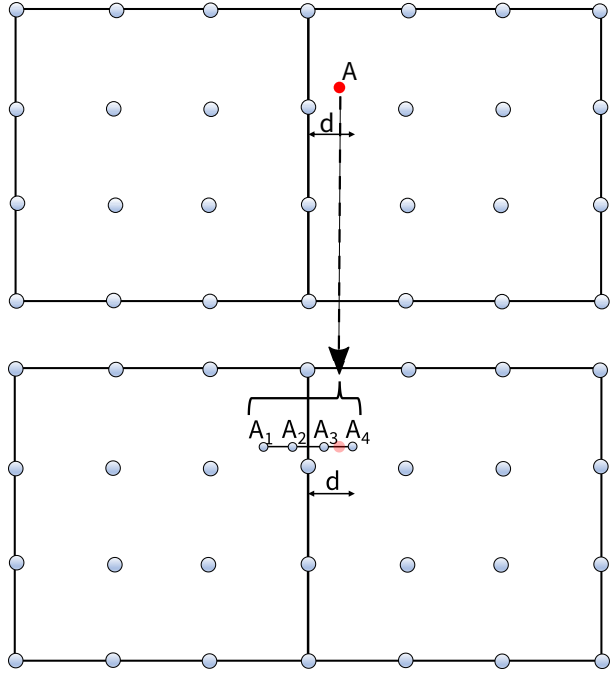}
\caption{Particles near a boundary of the finest cells.
The upper panel 1 shows the  a point charge near a boundary ($< d$), and the
lower panel 2 shows that the point charge is {\it replaced} ({\it split})
to $n$ point charges.
The squares are the finest cells and
the figure is drawn in two-dimension and for the case $n=4$ for simplicity.}
\label{arrow}
\end{center}
\end{figure}

\begin{figure}
\begin{center}
\includegraphics[width=8cm]{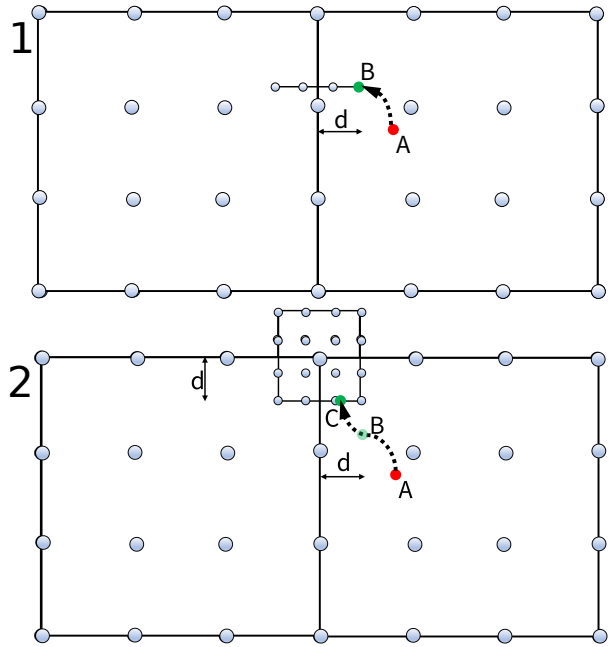}
\caption{Particles coming near boundaries of the finest cells.
The upper panel shows a point charge comes near a vertical boundary.
The moment the distance become to be $d$, it splits to $n$ point charges.
The lower panel shows the particle comes near another (horizontal) boundary.
When the distance from this horizontal boundary becomes $d$, it splits to $n^2$ point charges.
The figure is drawn in two-dimension for simplicity.}
\label{nearboundary}
\end{center}
\end{figure}
 
The idea to avoid this discontinuity is roughly as follows.
Since {\it replacement} replaces point charges by point charges fixed in position,
the motion of point charges cause the change
of strengths of fixed point charges, not the positions. 
Therefore, if we replace all the point charges and proceed as in the previous section,
we can avoid the discontinuity.

This scenario involves some problems.
One problem arises in the computation of near-field interaction.
In this approach, two distinct point charges may be in the same or adjacent cells, which causes
the potential between these two charges to go infinity and the method fails.
However, if
we further divide the finest cells so that no two particles are in the same cell and those cells containing particles 
are well-separated, we can avoid the problem.
It would be sufficient if the length of their edges $2d$ of the cells is (for example) $<0.1\AA$
because the intermolecular distances are $>1\AA$ in ordinary molecular simulations.
Denote the level of the further divided cells as $l_s$ ($l_s>l_f$ and $\frac{h}{2^{l_s}}\le 2d$).

In practice, in the computation of far-field interaction,
we do not need to replace most of the point charges with respect to the level $l_s$ cells.
If a point charge is in $C_{(l_s)}^i$ and the cell is in a finest cell $C_{(l_f)}^j$,
it is sufficient to {\it replace} the point directly to the cell $C_{(l_f)}^j$
because of the {\it invariance property}
$Rpl_{C_{(l_f)}^j}={\it shift}_{C_{(l_s)}^i\rightarrow C_{(l_f)}^j}\circ Rpl_{C_{(l_s)}^i}$.
Therefore, only point charges near boundaries of the finest cells 
need to be replaced with respect to their level $l_s$ cells in the 
far-field computation.

Even so,  in the near-field (direct) computation,
each point charge has to be {\it replaced} by $n^3$ point charges in its level $l_s$ cell
and computed directly between these points,
which is a severe burden.
So, in the direct computation,
we modify our method to replace (with respect to their level $l_s$ cells) only point charges
near boundaries of the finest cells as in the far-field computation.
This method also guarantees continuity.
In practice, our method is quite simple. We proceed as follows.
 
Firstly, suppose there is a point charge and 
there is one and only one boundary plane (of the finest cells) whose distance from the point charge is $< d$.
Also suppose $n$ is even.
Then, we {\it replace} it to $n$ point charges on a segment of length $2d$.
The segment is perpendicular to the boundary and its middle point is on the boundary.
Figure \ref{arrow} shows this for the case $n=4$.
The panel 1 depicts a point charge $A$ within a distance $d$ from a boundary and panel 2 depicts 
that it is replaced to point charges $A_1$, $A_2$, $A_3$, $A_4$.

If there are more than one boundaries near a given point, it is replaced to $n^2$ or $n^3$ points.
We describe this process using a moving point charge:
We consider a case where a point charge $A$ goes along the dotted curved arrow from $A$ to $C$ through $B$ as depicted in 
Fig.\ref{nearboundary}. 
When the charge reaches at the point $B$, where the distance from the left boundary is equal to $d$,
it is treated as $n$ points (panel 1 for $n=4$).
Then, this particle goes along the dotted curved arrow from $B$ to $C$ as depicted in the panel 2.
It has been treated as $4$ points from $B$ to just before $C$ and it is {\it replaced} to $4\times4$ points at $C$.
The figure is depicted in two-dimension for simplicity.
We understand in the same way for three-dimensional cases.
To distinguish this {\it replacement} from the {\it replacement} described in the previous sections, we call this {\it replacement}
as {\it split}.
Note that if the distance from a point is exactly $d$, the {\it split} charges are 0 except for the one that coincides with the point charge.

Now we have {\it split} point charges near boundaries ($<d$) and other point charges far from boundaries.
We take the new set of point charges again as given point charges in the simulation box, and
apply our method (both far-field and near-field as describes in \S\ref{formulation}).
Then we can avoid the discontinuity.
To compute force for a particle split near a boundary, we have to add
all the force the split particles feel.

So far, we have assumed that $n$ is even (for split). 
If $n$ is odd, we have to consider the assignment of
the middle particle on the segment. One way is to divide it into two same particles on the same position
(but the charge is half)
and assign them to two adjacent finest cells respectively. 
We could also slightly move the segment to make the middle point not to be on the boundary.

\section{Results}
\subsection{Accuracy and required time}
\label{accu1}
In this section, we investigate the accuracy of our {\it replacement} based method
and the time required to perform our method.
In our previous simulations of water,\cite{140} the finest cell was a cuboid with maximum
side-length less than 8.9 $\AA$. There were little more than 15 water molecules in the finest cells.
Fifteen water molecules have 45 atoms.
Therefore, in the following, we investigate accuracy and time required
for the cases  that there are 45 point charges in average in the finest cubic cells.
We first investigate the accuracy of potential energy and accuracy of force.

\begin{table}
\caption{Errors of potential and force and time required in computation.
The Error1 and Error2 are errors concerning to potential and force as described in \S\ref{accu1},
and Time required is the time elapsed to perform our FMM.
Times elapsed in the computation of near-field interaction are not included.
They were measured with respect to the order of replacement
on a Intel Core i7-9700 3GHz machine with intel Fortran compiler.
}
\label{t1}
\begin{center}
\begin{tabular}{clll}
\hline
\multicolumn{1}{c}{Order} & \multicolumn{1}{c}{Error1} & \multicolumn{1}{c}{Error2} & \multicolumn{1}{c}{Time required  (s)} \\
\hline
3 & 3.01E-2& 4.67E-2 & 2.05E-2\\
4 & 4.69E-3 & 8.20E-3 & 8.54E-2 \\
5 & 7.41E-4 & 1.47E-3 & 3.11E-1 \\
6 & 8.66E-5 & 2.87E-4 & 1.29 \\
7 & 1.36E-5 & 5.13E-5 & 3.40 \\
\hline
\end{tabular}
\end{center}
\end{table}

Table~\ref{t1} shows errors of potential energy and force, and average computation times required to compute the errors.
They were measured from order three to seven.
The computations were performed in a simulation box of cubic shape and the dimensions 
of the cube are all 1. The Coulomb potential $\frac{qq'}{r}$ was computed with respect to this length.
In the cube 23040 point charges were randomly scattered and their charge strengths are set to be $+1$ or $-1$
(see the bottom of this section).
The simulation box is divided into $8\times8\times8$ finest cells, which implies each 
finest cell contains 45 point charges in average.
In table~\ref{t1},
Error1 are errors of potential computed by $\frac{1}{n}\sum^n_i\left|\frac{{\rm PF}(i)-{\rm Direct}(i)}{{\rm Direct}(i)}\right|$
where $n$ is the number of particles ($=23040$), PF($i$) and Direct($i$) is the potential field that the
$i$-th particle feels computed by our method and direct computation, respectively.
The Error2 are errors of force defined by 
$\frac{1}{3n}\sum_i^n\sum_{j\in\{{x,y,z\}}} \left|\frac{{\rm FF}_j(i)-{\rm Direct}_j(i)}{{\rm Direct}_j(i)}\right|$
where ${\rm FF}_j(i)$ and ${\rm Direct}_j(i)$ are forces of $j$-direction ($j\in\{x,y,z\}$)
that $i$-th particle feels
computed by our method and direct computation, respectively.
As for the computation of ${\rm FF}_j(i)$, see the end of \S\ref{formulation}.
Time required is the average time elapsed to compute PF$(i)$ and FF${}_j(i)$.
Here, time required to compute near-field interaction is not included.
Computations were performed 10 times and the table shows their average time.

\begin{figure}
\begin{center}
\includegraphics[width=8cm]{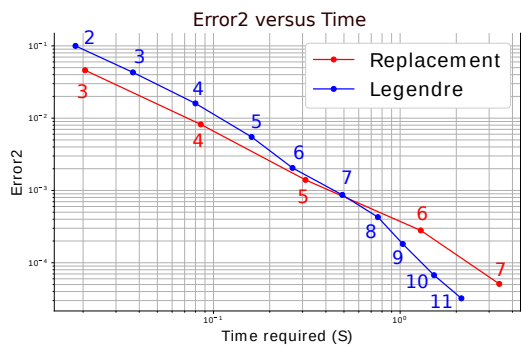}
\caption{Error2 versus time required, computed by our method and a method using Legendre's associated functions.
The numbers are orders of {\it replacement} and orders of multipoles.
Computation were performed for 23040 scattered point charges.
All charges are $+1$ or $-1$, and overall charge is 0.
Error2's were computed using the method described in the end of \S\ref{formulation}
and in the end of \S\ref{accu1}.
}
\label{EvT}
\end{center}
\end{figure}

Next we compare Error2 obtained by our method
with that by FMM that uses Legendre's associated functions.
For the latter case, to compute FF${}_j(i)$ which appear in the definition of Error2,
we made use of an existing program,
a sample program introduced in Ogata et al., \cite{ogata} 
setting the number of point charges = 23040 and
restricting it for one node. 
Figure \ref{EvT} shows {\it Error2 versus computation timings} for both methods,
and for various orders of {\it replacement} from three to seven (our method),
and various orders of multipoles from two to eleven (Legendre).

This sample program generates and scatters randomly given number of point charges. 
The half of them have charge strengths +1 and the other half $-1$.
Setting the number of point charges = 23040, we generated point charges and
used the same point charges
for all cases below and in the computation of errors above.

\subsection{Continuity}
\begin{figure}
\begin{center}
\includegraphics[width=8cm]{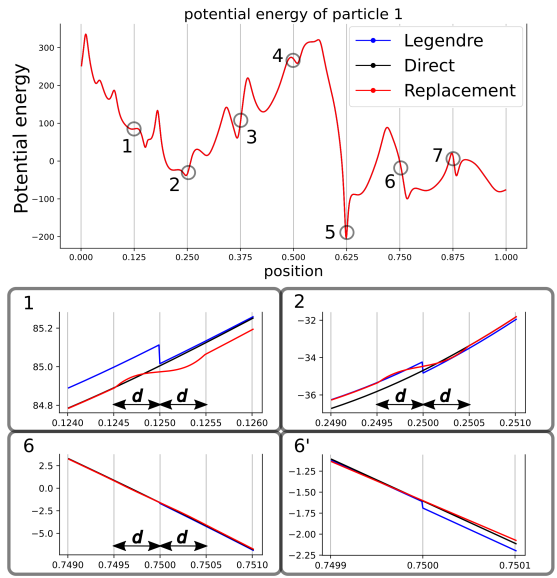}
\caption{Change of total potential energy with the motion of a point charge.
Three methods, using Legendre, direct computation, and ours (Replacement) are compared.
The order of multipole and {\it replacement} are both 4.
The upper panel indicates the change of the total potential energy with the motion of 
the point charge from one side of the simulation box to another along the direction of $x$-axis.
The vertical gray lines in the upper panel indicate the boundaries of the finest cells.
The numbered circles show the change of the potential energy near the boundaries.
The lower four panels magnifies the circles 1, 2, and 6.
The panel 6' is a further magnified  figure of the panel 6. Here $2d$ is set to $\frac{1}{1024}$.}
\label{Cont}
\end{center}
\end{figure}

The continuity of potential energy with respect to positions of 
point charges is obvious if we employ the {\it split} method (\S\ref{contBoun}).
Figure \ref{Cont} shows this continuity property.
It compares potential energy obtained from three methods, that is, using Legendre's associated functions (blue graph), 
direct computation (black graph), and our {\it replacement} based method (red graph).
The order of multipole for Legendre's and order of {\it replacement} are both set to 4.
In a simulation box, we prepared 23040 point charges described as before.
We used this box for the three cases.
We took the numbered one point charge in the sample program,\cite{ogata} and moved
it to 999 points successively from $x=\frac{1}{1000}$ to $x=\frac{999}{1000}$.
The $y$- and $z$-coordinate were not changed, i.e. they were the same as the original position.
The threshold $d$ was set to $2d=\frac{1}{1024}$.
Since each finest cell contains 45 charges in average, we estimate the length of the sides as about $8\AA$.
Thus the sides of the simulation box is $8\times8\AA=64\AA$.
If we choose $2d=0.1\AA$, it is $\frac{1}{640}$ of the length of the sides.
Thus we chose slightly smaller value $2d=\frac{1}{1024}$.

The upper panel of figure \ref{Cont} shows the change in total potential energy caused by the motion of number one point charge.
Each circle with a number illustrates the position where the number one charge pass through 
the boundaries.
We find that the three potential energy graphs almost overlap.
To scrutinize the graphs, we magnify these graphs $\times$200 near the boundaries.
The middle panels are magnified figures near circle 1 and circle 2.
We find graphs labeled as Legendre (blue graph) change discontinuously 
at the boundaries.
On the other hand, graphs of our method (labeled as Replacement, red graph) are continuous.
The graphs in other circles are too steep to see the sudden change, however,
if we further magnify the graph, we find the discontinuity of a blue graph and 
continuity of a red graph as illustrated in the bottom right panel for circle 6 (labeled as $6'$).

In this computation, 475 (or 476 when the moving point was replaced) point charges were {\it split}.
It caused about 10\% increase of computation time in both direct and far-field computation, and the errors
are almost the same (seemed slightly decreased).

\section{Discussion and conclusion}
\label{diss}

\subsection{Replacement}
We have introduced a {\it replacement} based FMM and investigated its accuracy and computational timings.
As shown in the table 1 and Fig.\ref{EvT}, our method performs similar to the existing 
Legendre's associated function based method.
Since our method uses neither complex numbers nor special functions, it is easy to implement.

\begin{figure}
\begin{center}
\includegraphics[width=8cm]{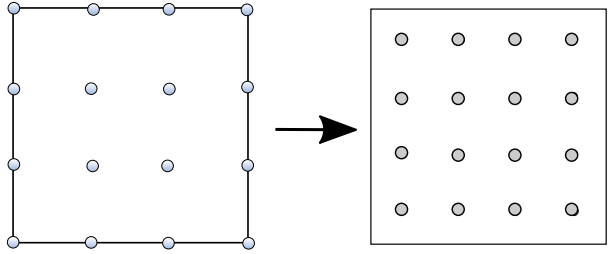}
\caption{$n$-division points in a cube.
The left panel shows the positions of fixed points we have been 
discussing (depicted for the case $n=4$ in two-dimension for simplicity).
The right panel shows slightly reduced arrangement of the points.
}
\label{Redu}
\end{center}
\end{figure}

The {\it replacement} method so far replaced point charges by point charges in
a cubic cell whose outermost
points are on the faces of the cube as in the Fig. \ref{F2}.
We can also take these $n$-division points in other locations such as the right square in Fig.\ref{Redu}, i.e.,
inside of the cube (Fig.\ref{Redu} is depicted in two-dimension for simplicity).
It was observed that when the distance of lattice points were reduced by $\times\frac{1}{1.2}$ in each dimension,
then error1 reduced about 20\% in  the case of order 4.
However, with the increase of orders, the improvement became insignificant, especially for orders $\ge 6$
even if we take other magnifications for the reduction.

There exists another interpolation based  FMM algorithm.
Our method is based on the Lagrange interpolation, however,
William Fong and Eric Darve developed a method using Chebyshev polynomials.\cite{120}
They placed more emphasis on approximation. 
On the other hand, our method is based on replacing arbitrary point charges by those on fixed positions.
Similar to their results, our method can be applied not only to the function $\frac{1}{|P-P_1|}$ but 
other functions such as $\frac{1}{|P-P_1|^2}$.
Therefore, we could directly compute forces by putting equations such as $\frac{x-x'}{|P-P_1|^3}$ instead of using
$\bm{v'}_a=-q\frac{\partial}{\partial a}g_{i,j,k}(x,y,z)|_{(x,y,z)=(x_1,y_1,z_1)}$
introduced in the end of \S\ref{formulation}.

\subsection{Continuity}
We have introduced a version of FMM which is continuous with the motion of particles
even if they go through the boundaries of the finest cells.
We would be able to apply this idea to periodic boundary conditions (PBCs).
We implement PBCs using Ewald summation or simply periodically pasting necessary
copies of the simulation box around the simulation box.
In both cases, potential energy and thus forces are not continuous on the boundaries of a simulation box in a strict sense.
The moment a point charge goes out from a simulation box, it appears from opposite side,
with a small discontinuity of potential energy.
We could apply our method to make them continuous at the boundaries.
We {\it split} point charges near boundaries of the simulation box.
Half of the $n$-division points (we assume $n$ is even for simplicity) are out side of the box.
We identify these outside points with the points in the simulation box 
that coincide with the outside points by adding or subtracting the length of the sides. 
In this way, we would be able to make the changes of potential energy continuous.

\appendix
\section{Appendix}
\subsection{Errors}
\label{AE11}
\begin{figure}
\begin{center}
\includegraphics[width=8cm]{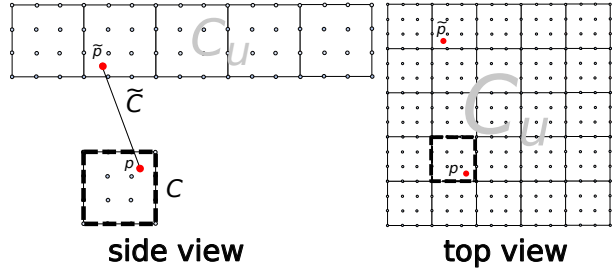}
\caption{The regions where $P$ and $\tilde{P}$ move. 
The left depicts the side view and the right depicts top view.
The left upper cuboid $C_u$ is composed of 25 cubes. 
These cubes are identical and the same as the cube below.
The edges of cubes are = 1. The distance from the cube containing $P$ to the cube right overhead is = 1.
The point $P$ moves in the cube $C$ with each step = 0.04, and the point $\tilde{P}$ moves in the 
cube $\tilde{C}$ in the upper cuboid (cubes) with each step = 0.04.
Then, $E_1$, the maximum of 
$E_1(C,\tilde{C})$,
is searched for all $\tilde{C}$ in the cuboid, and for a given order
(putting $q=\tilde{q}=1$).
}
\label{error3}
\end{center}
\end{figure}

In \S \ref{expansions}, we left the estimate of the error in the approximation equation (\ref{fmm}).
To estimate this error, we first consider the error of (\ref{newer}) as follows.
We take cubes $C$ and $\tilde{C}$ as in Fig.\ref{error3}.
The cube $C$ is the cube depicted below in the side view, and $\tilde{C}$ is one of the cubes (cuboid $C_u$) depicted above.
The cuboid $C_u$ is composed of 25 cubes as seen in the top view in Fig.\ref{error3}.
The dimensions of the cubes are all $1\times1\times1$.
Then, we consider $E_1:=\max_{\tilde{C}\in C_u} E_1(C,\tilde{C})$,
where $E_1(C,\tilde{C})$ is defined by
$E_1(C,\tilde{C}) :=\max_{P\in C, \tilde{P}\in\tilde{C}}\left| \frac{q\tilde{q}F(P,\tilde{P})-
\sum_{i,j,k}\sum_{\tilde{i},\tilde{j},\tilde{k}}g_{i,j,k}^{C}
g_{\tilde{i},\tilde{j},\tilde{k}}^{\tilde{C}}F(P(a_{i,j,k}),P(\tilde{a}_{\tilde{i},\tilde{j},\tilde{k}}))
}{ 
q\tilde{q}F(P,\tilde{P})}\right|.$
The numerator of equation above is the 
difference between the right-hand side and left-hand side of (\ref{newer}),
and the denominator is the left-hand side of (\ref{newer}).
In computing $E_1(C,\tilde{C})$,
we move $P$ and $\tilde{P}$ with the step 0.04 for $x$-, $y$-, and $z$-directions in the cube $C$ and $\tilde{C}$, respectively.
Thus $P$ and $\tilde{P}$ move on $26\times26\times26$ points.
Then we take the maximum of $E_1(C,\tilde{C})$ for all $\tilde{C}$ in the cuboid to obtain $E_1$.


\begin{table}
\caption{The maximum of $E_1$ and of $E_2$ 
with respect to the order of replacement.
Both errors are defined in Appendix\ref{AE11}.
$E_1$ is the error when both $P$ and $\tilde{P}$ are replaced,
 and $E_2$ is the error where $P$
is not replaced. 
}
\label{t2}
\begin{center}
\begin{tabular}{ccc}
\hline
\multicolumn{1}{c}{Order} & \multicolumn{1}{c}{$E_1$} & \multicolumn{1}{c}{$E_2$}  \\
\hline
3 & 3.57E-2 & 2.43E-2 \\
4 & 9.33E-3 & 6.92E-3  \\
5 & 3.20E-3 & 1.66E-3  \\
6 & 4.43E-4 & 3.69E-4  \\
7 & 1.04E-4 & 1.08E-4  \\
\hline
\end{tabular}
\end{center}
\end{table}

The table \ref{t2} shows {$E_1$} for order 3 to 7.
The table also shows $E_2$, which is defined by $E_2:=\max_{\tilde{C}\in C_u}E_2(C,\tilde{C})$, 
$
E_2(C,\tilde{C}):=\max_{P\in C, \tilde{P}\in\tilde{C}}\left| \frac{q\tilde{q}F(P,\tilde{P})-
\sum_{\tilde{i},\tilde{j},\tilde{k}}
g_{\tilde{i},\tilde{j},\tilde{k}}^{\tilde{C}}F(P,P(\tilde{a}_{\tilde{i},\tilde{j},\tilde{k}}))
}{ 
q\tilde{q}F(P,\tilde{P})}\right|.
$
The $E_2(C,\tilde{C})$ differs from $E_1(C,\tilde{C})$ in that $P$ in $E_2(C,\tilde{C})$ is not {\it replaced}.
At first glance, $E_1$ could be $n^3$ times as large as 
$E_2$, however, we find 
$\frac{E_1}{E_2}$ is at most 2.
In all cases of orders, $E_1(C,\tilde{C})$ and $E_2(C,\tilde{C})$ took its maximum when $\tilde{C}$ was just above $C$.
In addition, if we increase the distance between $C$ and a cell $\tilde{C}$ just above, 
the error $E_1(C,\tilde{C})$ decreased rapidly.
Since the positional relations depicted in Fig.\ref{error3} essentially exhaust those appear in 
the interaction list (sixteen cubes may be sufficient, but the computation was performed for these 25 cubes),
the error $E_1$ gives us an estimate of error in using our method.

Lastly, we consider the error of (\ref{fmm}).
Since $E_1\le \alpha$ for a number $\alpha$ implies 

\[
\left|\sum_{(P,q)\in Pc}\sum_{(\tilde{P},\tilde{q})\in P\tilde{c}}q\tilde{q}F(P,\tilde{P})-
\sum_{i,j,k}\sum_{\tilde{i},\tilde{j},\tilde{k}}g_{i,j,k}^{C}
g_{\tilde{i},\tilde{j},\tilde{k}}^{\tilde{C}}F(P(a_{i,j,k}),P(\tilde{a}_{\tilde{i},\tilde{j},\tilde{k}}))\right|
\le \alpha|q\tilde{q}F(P,\tilde{P})|,
\]
the absolute value of the error of (\ref{fmm}) is
$\le\alpha\sum_{l,\tilde{l}}|q_l\tilde{q}_l|\times \max F(P,\tilde{P})
=\alpha(\sum_{l}|q_l|)(\sum_{\tilde{l}}|\tilde{q}_l|)\times \max F(P,\tilde{P})$
(in the case above, $\max F(P,\tilde{P})=1$).
The number $\alpha$ is given as $E_1$ in the table  \ref{t2}.

\subsection{Bilinear forms - removing shift process}
\label{appBi}
We can remove {\it shift} processes from our method.
In our {\it replacement} based method, we {\it shift} {\it replaced} vectors downward and upward 
to compute $B_{C,\tilde{C}}^n(\Phi_C,\Phi_{\tilde{C}})$,
which gives us an approximation of interactions between the point charges in $C$ and those in $\tilde{C}$.
Here, $\Phi_C$ is a vector obtained by replacing the point charges in $C$ {\it with respect to $C$} as
defined in \ref{formulation}.
Even if $C$ is composed of eight children $C_i$ ($C=\bigsqcup_{1\le i\le8}C_i$ (disjoint union)), 
$\Phi_{C}\ne\sum_{1\le i\le8}\Phi_{C_i}$ and thus
$B_{C,\tilde{C}}^n(\sum_i\Phi_{C_i},\Phi_{\tilde{C}})\ne B_{C,\tilde{C}}^n(\Phi_C,\Phi_{\tilde{C}})$.
However, $\Phi_{C}=\sum_{1\le i\le8}{\it shift}_{C_i\rightarrow C}(\Phi_{C_i})$.
Therefore we have repeatedly {\it shifted} vectors.

Nevertheless, we can remove this {\it shift} process as follows.
Let $\boldsymbol{S}$ be a cubic simulation box.
Since
${\it shift}_{\boldsymbol{S}\rightarrow C}({\it shift}_{C\rightarrow \boldsymbol{S}}(\bm{v}))=\bm{v}$
for any vector $\bm{v}$, we have
\begin{equation}
B_{C,\tilde{C}}^n(\bm{v},\bm{w})
=B_{C,\tilde{C}}^n({\it shift}_{\boldsymbol{S}\rightarrow C}({\it shift}_{C\rightarrow \boldsymbol{S}}(\bm{v})),
{\it shift}_{\boldsymbol{S}\rightarrow \tilde{C}}({\it shift}_{\tilde{C}\rightarrow \boldsymbol{S}}(\bm{w}))).
\end{equation}
Thus, putting 
${\mathcal B}_{C,\tilde{C}}^n(\bm{v},\bm{w}):=
B_{C,\tilde{C}}^n({\it shift}_{\boldsymbol{S}\rightarrow C}(\bm{v}),{\it shift}_{\boldsymbol{S}\rightarrow \tilde{C}}(\bm{w}))$
and $\Phi_{C}^{\boldsymbol{S}}:={\it shift}_{C\rightarrow \boldsymbol{S}}(\Phi_C)$,
we have
\begin{equation}
B_{C,\tilde{C}}^n(\Phi_C,\Phi_{\tilde{C}})
={\mathcal B}_{C,\tilde{C}}^n(\Phi_C^{\boldsymbol{S}},\Phi_{\tilde{C}}^{\boldsymbol{S}}).
\end{equation}
Here, 
$\Phi_{C}^{\boldsymbol{S}}:={\it shift}_{C\rightarrow \boldsymbol{S}}(\Phi_C)$ is equal to a vector
obtained by directly replacing the point charges in $C$ with respect to $\boldsymbol{S}$.
Therefore, if $C=\bigsqcup_{1\le i\le8}C_i$, $\sum_i\Phi_{C_i}^{\boldsymbol{S}}=\Phi_{C}^{\boldsymbol{S}}$ and thus
${\mathcal B}_{C,\tilde{C}}^n(\sum_i\Phi_{C_i}^{\boldsymbol{S}},\Phi_{\tilde{C}}^{\boldsymbol{S}})
=B_{C,\tilde{C}}^n(\Phi_C,\Phi_{\tilde{C}})$.
This makes Upward pass a simple addition of vectors,
and downward pass is identity (do not change).
What we need to do is to replace the point charges in a finest cell with respect to $\boldsymbol{S}$ and compute
${\mathcal B}_{C,\tilde{C}}$.
Therefore, if we compute the bilinear forms ${\mathcal B}_{C,\tilde{C}}$ 
for all $C$ and $\tilde{C}$
and save them in the RAM in advance, it is likely to save computation time.
A preprint\cite{arX} has performed this FMM for order = 4, and it performed a little faster than the sample program.\cite{ogata}
An issue that arises from saving all
${\mathcal B}_{C,\tilde{C}}^n$ in memory is that it consumes much memory.
Therefore, saving all ${\mathcal B}_{C,\tilde{C}}^n$ is feasible 
only when $l_f$ is small. In the preprint, simulations were performed for $l_f = 3$.




\end{document}